\documentclass[useAMS,usenatbib]{mn2e}

\usepackage{epsfig}
\usepackage{graphics}

\newcommand{\Lsun}{\mbox{$\rm L_{\odot}$}}

\def\aFe{[$\alpha/{\rm Fe}$]}
\def\Hb{${\rm H}{\beta}$}
\def\Mgb{{\rm Mg}$b$}
\def\Fe{$\langle {\rm Fe}\rangle$}
\def\FeH{[{\rm Fe}$/${\rm H}]}
\def\MgFe{[${\rm MgFe}$]$'$}

\title[The nuclear stellar disc of NGC~1023]{The young nuclear stellar disc in the SB0 galaxy NGC~1023}

\author[E. M. Corsini et al.]{E. M. Corsini,$^{1,2,}$\thanks{E-mail:
enricomaria.corsini@unipd.it} 
L. Morelli,$^{1,2}$ N. Pastorello,$^3$ 
E. Dalla Bont\`a,$^{1,2}$ A. Pizzella$^{1,2}$ \and  
and E. Portaluri$^{2}$\\
\\$^{1}$Dipartimento di Fisica e Astronomia `G. Galilei', 
 Universit\`a di Padova, vicolo dell'Osservatorio 3, I-35122 Padova, Italy
\\$^{2}$INAF--Osservatorio Astronomico di Padova, vicolo
 dell'Osservatorio 5, I-35122 Padova, Italy
\\$^3$Centre for Astrophysics and Supercomputing, 
Swinburne University of Technology, Hawthorn, VIC 3122, Australia}

\begin{document}

\date{\today}

\pagerange{\pageref{firstpage}--\pageref{lastpage}} \pubyear{0000}

\maketitle

\label{firstpage}

\begin{abstract}
Small kinematically decoupled stellar discs with scalelengths of a few
tens of parsec are known to reside in the centre of
galaxies. Different mechanisms have been proposed to explain how they
form, including gas dissipation and merging of globular
clusters. Using archival {\em Hubble Space Telescope} imaging and
ground-based integral-field spectroscopy, we investigated the
structure and stellar populations of the nuclear stellar disc hosted
in the interacting SB0 galaxy NGC 1023. The stars of the nuclear disc
are remarkably younger and more metal rich with respect to the host
bulge. These findings support a scenario in which the nuclear disc is
the end result of star formation in metal enriched gas piled up in the
galaxy centre. The gas can be of either internal or external origin,
i.e. from either the main disc of NGC~1023 or the nearby satellite
galaxy NGC~1023A. The dissipationless formation of the nuclear disc
from already formed stars, through the migration and accretion of star
clusters into the galactic centre, is rejected.
\end{abstract}


\section{Introduction}
\label{sec:introduction}

Small discs of stars with scalelengths of a few tens of parsecs and
luminosities up to $10^7$ \Lsun\ are known to reside in the nuclei of
galaxies \citep{Pizzella2002, Ledo2010}.

Different mechanisms have been proposed to explain how nuclear stellar
discs (NSDs) assembled.
Most of them are believed to be formed from gas funnelled into the
nucleus, either via bar-driven secular infall
\citep[][]{Silchenko1997, Scorza1998, Krajnovic2004} or by external
accretion \citep[][]{Pizzella2002, Morelli2004, Corsini2012}. In both
scenarios, the gas is efficiently driven toward the galactic centre,
where first it settles as it dissipates and then it turns into stars
as density rises. Ionized gas and dust are actually observed in a few
nuclear discs \citep{Kormendy1994, Kormendy2005}.

This picture has been recently challenged by the simulations analysed
by \citet{Portaluri2013}. They found that most of the mass of NSDs can
be assembled from already formed stars through the migration and
accretion of star clusters into the galactic centre. However, some gas
is needed in the very early stages of the formation of the flattened
NSD. Moreover, the subsequent dissipationless building of the NSD
works only for a very narrow range of initial conditions.

In this context, NSDs could be a powerful tool to constrain the
assembly history of galaxies. This is because recent numerical
simulations have shown that NSDs are indeed fragile structures that
are easily disrupted during a major merger event
\citep{Sarzi2015a}. Therefore, the stellar age of NSDs may be adopted
as a proxy to date the epoch since their host galaxies experienced the
last major merging.

To date, the properties of the stellar populations (i.e. age,
metallicity, and star formation time scale) in the nuclear regions of a
number of galaxies hosting an NSD have been studied in some detail
using colour-colour diagrams \citep{vandenBosch_etal1998},
line-strength indices \citep{Silchenko1999, Morelli2004,
  Krajnovic2004}, and population synthesis models
\citep{Sarzi2005}. However, all these results concern the total
stellar population of the galactic nucleus without properly separating
the contribution of the NSD from that of the host galaxy. The only
exception is the elliptical galaxy NGC~4458, where \citet{Sarzi2015b}
found that the NSD is at least 6 Gyr old. To precisely constrain the
stellar population of an NSD it is indeed necessary to perform a
detailed measurement of its light distribution. This is quite
challenging since the luminosity of NSDs is only a few per cent of
that of the galactic nucleus and locally NSDs contribute at most half
of the total surface brightness \citep{Pizzella2002, Morelli2004,
  Morelli2010, Corsini2012}.

These findings call for a systematic investigation of the NSD stellar
ages in galaxies of different mass and across different environments
in order to time their assembly. This will allow to test some of the
predictions for the accretion phase in the two-phase galaxy formation
scenario \citep[e.g.][]{Naab2007, Oser2010, Lackner2012}. For
instance, older NSDs are expected to reside in less massive cluster
galaxies. Indeed, they should have assembled earlier and have
experienced less merging events, capable to destroy the nuclear disc,
than their counterparts in the field.

In this paper, we revisit the case of NGC~1023 since it represents an
excellent test case to derive the properties of the NSD stellar
population.
NGC~1023 is a highly-inclined disc galaxy classified as SB0$_1$(5) by
\citet{RSA} and SB0$^-$(rs) by \citet[][hereafter RC3]{RC3}. Its total
$B$-band magnitude is $B_T=10.35$ (RC3), which, after correcting for
inclination and extinction, corresponds to $M_{B_T}^{0} = -20.11$ for
an adopted distance of $10.9$ Mpc \citep[][$H_0=75$
 km~s~Mpc$^{-1}$]{Faber1997}.
NGC~1023 is the brightest member of the LGG 70 group
\citep{Garcia1993}. Its closest companion, NGC~1023A, is a Magellanic
irregular dwarf galaxy of low surface brightness \citep{CAG}, located
about $2\farcm7$ (RC3) East of NGC~1023 centre. This corresponds to a
projected linear distance of about $8$ kpc. The two galaxies are
connected by a bridge of neutral hydrogen which suggests an on-going
interaction \citep{Sancisi1984}. The global properties and spectral
energy distribution of NGC~1023 are suggestive of a minor merger
occurred $\sim2$ Gyr ago \citep{Bettoni2012}.
The broad-band images of the centre of NGC~1023 obtained by {\em
  Hubble Space Telescope (HST)\/} show an almost edge-on NSD
\citep{Faber1997, Silchenko1999, Sarzi2006, Ledo2010}. The stellar
dynamics confirms the presence of a nuclear flattened component, as it
results from the tangentially anisotropic distribution of the
innermost stellar orbits \citep{Bower2001}. Finally, the stellar
population of the nucleus shows significantly different chemical
properties from the rest of the galaxy. In particular, the nucleus is
younger, more metal rich, and it shows higher magnesium overabundance
than the surrounding bulge \citep{Silchenko1999, McDermid2006}.

Here, we improve the previous results by quantitatively
constraining the light distribution, stellar age, and iron abundance
of the NSD of NGC~1023.
The photometric decomposition of archival multiband optical images of
the galaxy nucleus obtained with {\em HST} allows us to get the basic
structural parameters of the NSD (central surface brightness,
scalelength radius, inclination, and major-axis position
angle) in Section~\ref{sec:structure}. 
The contribution of the NSD to the total surface-brightness
distribution is adopted in Section~\ref{sec:population} in combination with
stellar population models to derive at last the actual properties of
the stellar population (age and iron abundance) of the NSD and its
host bulge from the line-strength indices measured in the galaxy
nucleus.
Our findings about the formation process of the NSD are discussed in
Section~\ref{sec:discussion}.

\section{Photometric properties of the nuclear disc}
\label{sec:structure}

\subsection{Broad-band imaging}
\label{sec:imaging}

We retrieved from the {\em HST} Science Data Archive the images of
NGC~1023 obtained with the Advanced Camera for Survey (ACS) and filter
{\em F475W} (Prop. Id. 12202, PI: G. Sivakoff) and with the Wide Field
Planetary Camera 2 (WFPC2) and filters {\em F555W} and {\em F814W}
(Prop. Id. 6099, PI: S. M. Faber). These three data sets were selected
as a compromise to analyse the deepest and unsaturated broad-band
images of the galaxy nucleus obtained at the highest spatial
resolution.

The ACS images were taken with the Wide Field Channel (WFC), which
consists of two SITe CCDs with $2048\,\times\,4096$ pixels each of
size $15\,\times\,15$ $\mu$m$^2$. The image scale is $0.049$ arcsec
pixel$^{-1}$ and the field of view of the combined detectors covers an
approximately square area of about $202\,\times\,202$ arcsec$^2$. The
gain and readout noise of the four WFC amplifiers are about 2.0 $e^-$
count$^{-1}$ and 4.2 $e^-$ (rms), respectively.
All the WFPC2 exposures were taken by centring the galaxy nucleus on
the Planetary Camera (PC). The PC detector is a Loral CCD with
$800\,\times\,800$ pixels and a pixel size of $15\,\times\,15$
$\mu$m$^2$. The image scale of 0.046 arcsec pixel$^{-1}$ yields a
field of view of about $36\,\times\,36$ arcsec$^2$. The gain and
readout noise are 14.0 $e^-$ count$^{-1}$ and 7.0 $e^-$ (rms),
respectively.
To help in identifying and correcting cosmic ray events, different
exposures were taken with each filter.  The total exposure time was
1552 s for the {\em F475W} filter, 1620 s for the {\em F555W} filter,
and 1880 s for the {\em F814W} filter. For both ACS and WFPC2 images
the telescope was always guided in fine lock, giving a typical rms
tracking error per exposure of 0.005 arcsec.

The ACS images were calibrated using the {\sc calacs} reduction
pipeline in {\sc iraf}\footnote{The Imaging Reduction and Analysis
  Facility ({\sc iraf}) is distributed by the National Optical
  Astronomy Observatory, which is operated by the Association of
  Universities for Research in Astronomy (AURA), Inc., under
  cooperative agreement with the National Science
  Foundation.}. Reduction steps include bias subtraction, dark current
subtraction, flat-fielding correction, and correction for geometric
distortion with {\sc iraf} task {\sc multidrizzle}
\citep{Fruchter2009} as described in detail in ACS instrument and data
handbooks \citep{Pavlovsky2004, Pavlovsky2006}. The images were
aligned by comparing the centroids of stars in the field of view and
then combined, rejecting cosmic rays in the process.  Residual cosmic
ray events and hot pixels were removed using the {\sc lacos\_ima}
procedure \citep{vanDokkum2001}.  The sky level was determined from
regions free of sources at the edge of the field of view and then
subtracted.
The WFPC2 images were reduced using the {\sc calwfpc} reduction
pipeline in {\sc iraf}. Reduction steps include bias subtraction, dark
current subtraction, and flat-fielding, as described in detail in the
WFPC2 instrument and data handbooks \citep{Baggett2002,
 McMaster2008}. Subsequent analysis including alignment and
combination of images and rejection of cosmic rays was performed using
{\sc multidrizzle} and {\sc lacos\_ima}. The sky level was determined
from regions free of sources in the Wide Field (WF) chips and
subtracted from the combined PC frames after appropriate scaling.

The ACS/{\em F475W}, WFPC2/{\em F555W}, and WFPC2/{\em F814W}
passbands are similar to Johnson-Cousins $B$, $V$, and $I$ bands,
respectively. The flux calibration to the Vega magnitude system in the
three observed {\em HST\/} passbands was performed following
\citet{Holtzman1995}, \citet{Sirianni2005}, and \citet{Bohlin2012}, and
included aperture and gain corrections.

\subsection{Photometric model}
\label{sec:decomposition}

The surface-brightness distribution of the NSD of NGC~1023 was
independently derived in the different passbands using the method of
\citet{Scorza1995} as implemented by \citet{Morelli2004}. But, we
adopted a different best-fitting algorithm to perform the photometric
decomposition and followed the prescriptions of \citet{Morelli2010}
and \citet{Corsini2012} for the treatment of the isophotal shape of
the bulge and to account for the point-spread function (PSF),
respectively.
The photometric decomposition is based on the assumption that the
isophotal disciness, quantified by the positive value of the
fourth cosine Fourier coefficient $A_4$ \citep{Jedrzejewski1987,
  Bender1988}, is the result of the superimposition of the light
contribution of a rounder host bulge and a more elongated nuclear
disc.
We assumed that the nuclear disc has perfectly elliptical isophotes
($A_{\rm 4,NSD} = 0$) with constant ellipticity, $\epsilon_{\rm NSD}$,
and constant position angle, PA$_{\rm NSD}$, whereas the bulge
has elliptical isophotes with constant $A_{\rm 4,bulge}$,
constant ellipticity $\epsilon_{\rm bulge}$, and constant position
angle PA$_{\rm bulge}$. Moreover, the bulge isophotes can be
either perfectly elliptical ($A_{\rm 4,bulge} = 0$) or be
characterised by a constant discy ($A_{\rm 4,bulge} > 0$) or boxy
shape ($A_{\rm 4,bulge} < 0$). 

Let $(x,y)$ be the Cartesian coordinates in sky plane with the
origin in the galaxy centre, the $x$-axis parallel to the direction of
right ascension and pointing westward, the $y$-axis parallel to the
direction of declination and pointing northward.
We assumed the surface brightness of the NSD to follow the exponential
law \citep{Freeman1970} and have elliptical isophotes centred on
$(x_0,\,y_0)$ with constant PA$_{\rm NSD}$ and constant
$\epsilon_{\rm NSD} = 1 - q_{\rm NSD}$ where $q_{\rm NSD}$
is the isophotal minor-to-major axis ratio.  We considered the disc to
be infinitesimally thin and derived its inclination as $i =
\arccos{(q_{\rm NSD})}$.  The disc surface-brightness distribution is
given by:

\begin{equation} 
I(x,y) = I_0\,\exp{\left[-\frac{r(x,y)}{h}\right]}, 
\label{eqn:disc_sb} 
\end{equation} 
%
where $I_0$ and $h$ are the central surface brightness and
scalelength of the disc, respectively, and $r$ is:

\begin{eqnarray} 
r(x,y) & = & \left[(-\Delta x \sin{{\rm PA_{\rm NSD}}} 
                   +\Delta y \cos{{\rm PA_{\rm NSD}}})^2 - \right. \nonumber \\  
       &   & \hspace{-0.5cm}\left.(\Delta x \cos{{\rm PA_{\rm NSD}}} 
                   -\Delta y \sin{{\rm PA_{\rm NSD}}})^2/q_{\rm NSD}^2\right]^{1/2}, 
\label{eqn:disc_radius} 
\end{eqnarray} 
%
where $\Delta x = x-x_0$ and $\Delta y = y-y_0$.

\begin{figure*}
\includegraphics[width=14cm,angle=0]{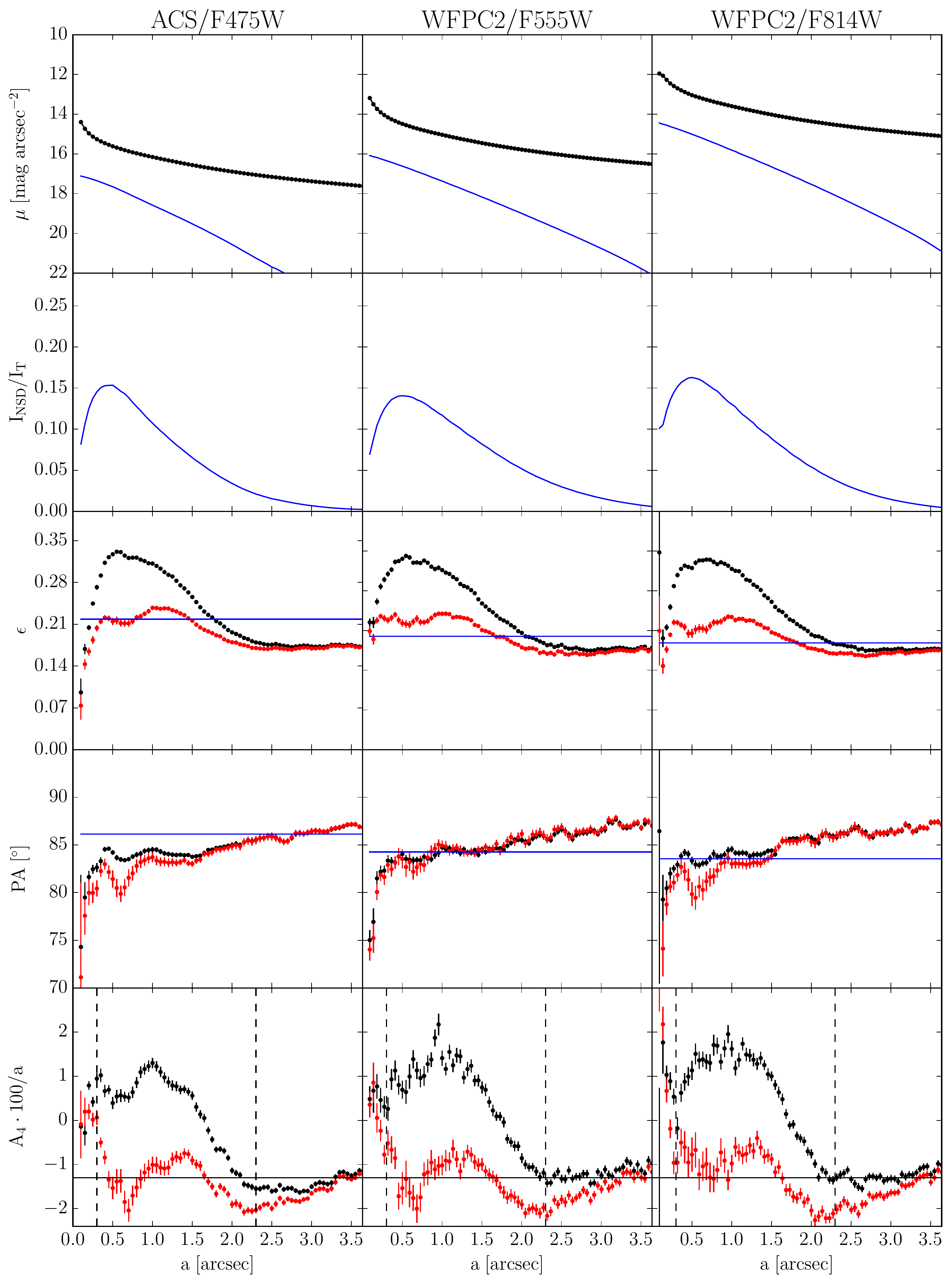}
\caption{Isophotal parameters of the nuclear region of NGC~1023 as a
  function of the isophotal semi-major axis based on the analysis of
  the surface-brightness distribution measured in the ACS/{\em F475W}
  (left-hand panels), WFPC2/{\em F555W} (central panels), and
  WFPC2/{\em F814W} (right-hand panels) images, respectively. From top
  to bottom: surface-brightness radial profiles of the galaxy (filled
  black circles) and NSD (after convolution with the {\em HST} PSF,
  dotted blue line); radial profile of the NSD-to-total
  surface-brightness ratio; radial profiles of the galaxy ellipticity,
  position angle, and fourth cosine Fourier coefficient before (filled
  black circles) and after (open red circles) the subtraction of the
  best-fitting model for the NSD. For each image the best-fitting
  ellipticity $\epsilon_{\rm NSD}$ and position angle PA $_{\rm NSD}$
  of the NSD are marked with dotted blue lines. The vertical dashed
  lines bracket the radial range of $A_4$ where $\chi^2$ was computed
  while a solid black line marks the adopted value of $A_{\rm
    4,bulge}$.}
 \label{fig:decomposition}
\end{figure*}

The isophote fitting with ellipses was carried out on the galaxy image
using the {\sc iraf} task {\sc ellipse}
\citep{Jedrzejewski1987}. First, isophotes were fitted by ellipses
allowing their centres to vary. Within the errors, no variation in the
ellipse centres was found. Therefore, we assumed the disc centre to be
coincident with the galaxy centre. The final ellipse fits were done at
fixed ellipse centres out to a distance of 3.6 arcsec.
The ellipse-averaged profiles of the surface brightness, position
angle, ellipticity, and fourth cosine Fourier coefficient are plotted
in Fig.~\ref{fig:decomposition}.

\begin{figure*}
\includegraphics[width=14cm,angle=0]{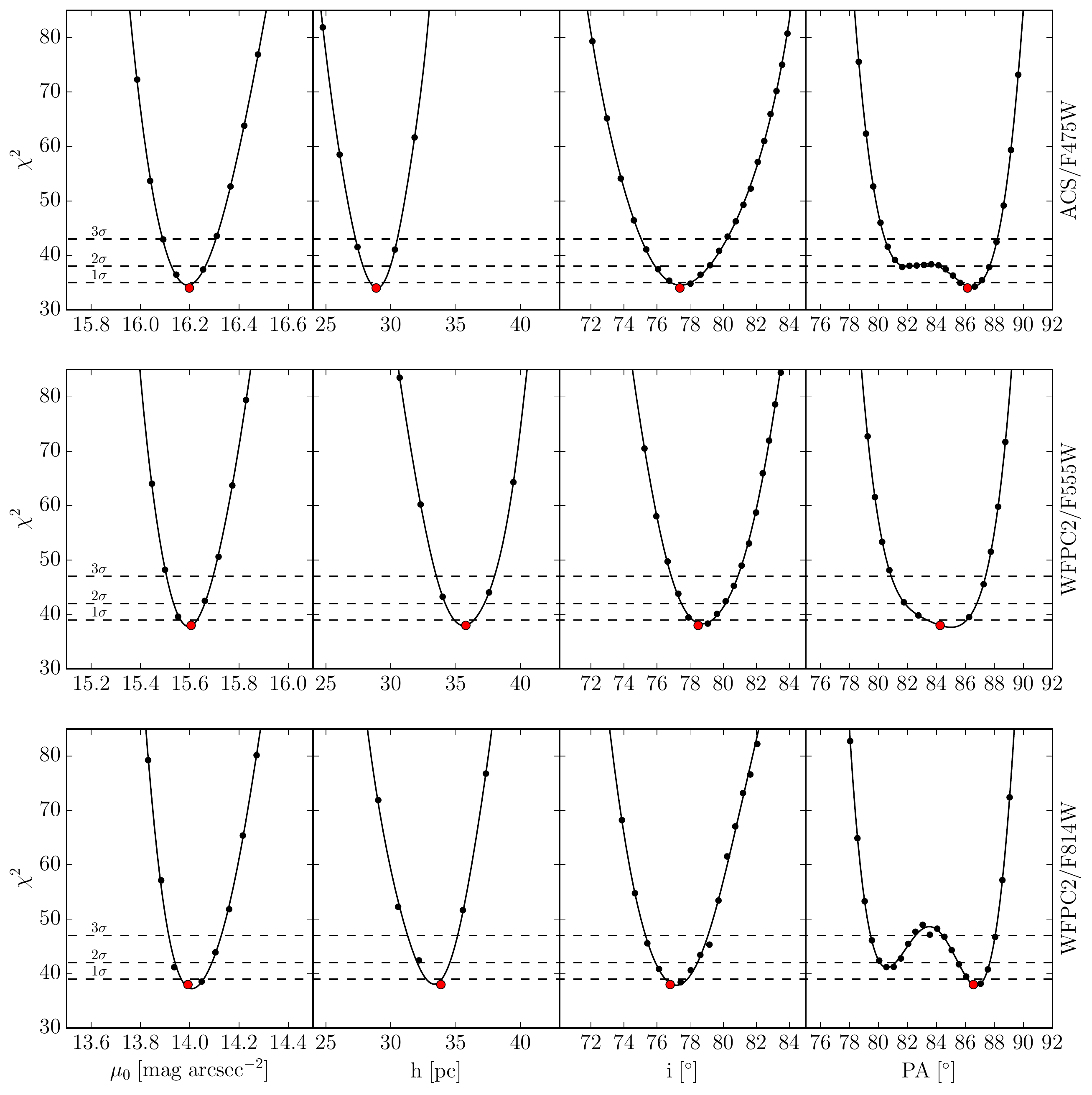}
\caption{$\chi^2$ distribution for the photometric decomposition of
  the ACS/{\em F475W} (top panels), WFPC2/{\em F555W} (middle panels),
  and WFPC2/{\em F814W} (bottom panels) images of NGC~1023 as a
  function of $\mu_0$, $h$, $i$, and PA (from left to right),
  respectively. The dotted horizontal lines indicate the 1$\sigma$,
  2$\sigma$, and 3$\sigma$ confidence levels on the best-fitting
  values, marginalizing over all other parameters.}
 \label{fig:confidence}
\end{figure*}

We derived the photometric parameters of the nuclear disc ($I_0$, $h$,
$q_{\rm NSD}$, and PA$_{\rm NSD}$) by iterative subtraction of the
model surface brightness given by Eq.~\ref{eqn:disc_sb} from the
observed surface-brightness distribution of the galaxy. 
For each nuclear disc model, the disc-free image of the galaxy was
obtained from the galaxy image by subtracting the nuclear disc model
after convolving with the {\em HST} PSF. The adopted PSF model was
calculated with the {\sc tinytim} package \citep{Krist1999} taking
into account the instrumental set-up and position of the NSD on the
given image.

Then, we performed an isophotal analysis of the disc-free image using
{\sc ellipse}. The nuclear disc parameters were adjusted until the
isophotes in the disc-free image have $A_{\rm 4,disc-free} = A_{\rm
  4,bulge}$ out to the very centre. For the bulge of NGC~1023, we
derived the mean $A_4$ between 2.3 and 3.3 arcsec outside the region
where the inner peak of $A_4$ is observed. The mean values measured in
the different passbands are consistent within the errors. Therefore,
we calculated $A_{\rm 4,bulge} \pm \sigma_{A{\rm 4,bulge}}
=-(13.5\pm0.2)\times10^{-3}$ as the mean and standard deviation of the
mean of all the available $A_4$ values between 2.3 and 3.3 arcsec,
respectively.
We did no assumption for the radial profiles of surface brightness,
ellipticity, and position angle of the bulge which is triaxial
\citep{Silchenko2005}.
We calculated:
\begin{equation}
\chi^2 = \sum_{i=1}^N \frac{(A_{{\rm 4,disc-free,}i}-A_{\rm
    4,bulge})^2}{\sigma^2_i+\sigma^2_{A{\rm 4,bulge}}}
\end{equation}
where $A_{{\rm 4,disc-free},i}\pm\sigma_i$ is the value of the $A_4$
Fourier coefficient measured for the $i$-th isophote in the disc-free
image and $N$ is the number of fitted isophotes in the region of the
NSD. This is the radial range between 0.3 and 2.3 arcsec, which is
bracketed by two vertical lines in Fig.~\ref{fig:decomposition}. We
assumed $\sigma_i$ as the mean error $\langle\sigma\rangle$ on the
values of $A_{\rm 4,disc-free}$ in the NSD region of the
best-fitting disc-free image. It is $\langle\sigma\rangle =
1.5\times10^{-3}, 2.1\times10^{-3}$, and $2.4\times10^{-3}$ for the
ACS/{\em F475W}, WFPC2/{\em F555W}, and WFPC2/{\em F814W} images,
respectively.
The disc models resulting in $| \langle A_{\rm 4,disc-free} \rangle\,
- A_{\rm 4,bulge} |>\,0.003$ were rejected to avoid unrealistic
solutions.

\begin{table*}
\caption{Photometric parameters of the nuclear stellar disc.}
\begin{tabular}{lcccccc}
\hline
\multicolumn{1}{c}{Filter} &
\multicolumn{1}{c}{$\mu_0$} &
\multicolumn{1}{c}{$h$} &
\multicolumn{1}{c}{$i$} &
\multicolumn{1}{c}{PA$_{\rm NSD}$} &
\multicolumn{1}{c}{$L_{\rm NSD}$}\\
\multicolumn{1}{c}{} &
\multicolumn{1}{c}{[mag arcsec$^{-2}$]} &
\multicolumn{1}{c}{[pc]} &
\multicolumn{1}{c}{[$^\circ$]} &
\multicolumn{1}{c}{[$^\circ$]} &
\multicolumn{1}{c}{[$10^7\,{\rm L}_\odot$]} \\
\multicolumn{1}{c}{(1)} &
\multicolumn{1}{c}{(2)} &
\multicolumn{1}{c}{(3)} &
\multicolumn{1}{c}{(4)} &
\multicolumn{1}{c}{(5)} &
\multicolumn{1}{c}{(6)} \\
\hline
ACS/{\em F475W}   & $16.20^{+0.11}_{-0.11}$ & $28.9^{+1.6}_{-1.6}$  & $77.4^{+2.8}_{-2.3}$ & $86.1^{+2.1}_{-5.7}$ & $1.9^{+0.9}_{-0.7}$\\
WFPC2/{\em F555W} & $15.61^{+0.09}_{-0.10}$ & $35.8^{+2.2}_{-2.3}$  & $78.5^{+2.4}_{-1.6}$ & $84.3^{+3.1}_{-3.3}$ & $3.3^{+1.3}_{-1.2}$\\
WFPC2/{\em F814W} & $13.99^{+0.13}_{-0.08}$ & $33.9^{+1.4}_{-2.6}$  & $76.8^{+2.3}_{-1.5}$ & $86.6^{+1.5}_{-2.1}$ & $7.8^{+2.3}_{-2.9}$\\
\hline
\end{tabular}
\begin{minipage}{11cm}
{\em Note.} Column (1): passband.  Column (2): observed central
surface brightness. Column (3): scalelength. Column (4): inclination
derived as $i = \arccos{(q_{\rm NSD})}$. Column (5): major-axis position
angle. Column (6): total luminosity derived as $L_{\rm NSD}\,=\,2
\pi I_0 h^2 q_{\rm NSD}$.
\end{minipage}

\label{tab:parameters}
\end{table*}

We explored the parameter space in a twofold process by running a
{\sc pyraf}\footnote{{\sc pyraf} is a product of the Science Software Branch at
 the Space Telescope Science Institute.} code. First, we built a set
of NSD models covering the central surface brightnesses
$15.7\,\leq\,\mu_{0,{\rm F475W}}\,\leq\,17.8$ mag~arcsec$^{-2}$,
$14.1\,\leq\,\mu_{0,{\rm F555W}}\,\leq\,16.2$ mag~arcsec$^{-2}$,
$13.0\,\leq\,\mu_{0,{\rm F814W}}\,\leq\,15.1$ mag~arcsec$^{-2}$,
scalelengths
$8\,\leq\,h\,\leq\,65$ pc, 
axial ratios 
$0.04\,\leq\,q_{\rm NSD}\,\leq\,0.35$,
and position angles
$76^\circ\,\leq\,{\rm PA_{\rm NSD}}\,\leq\,92^\circ$.
The model corresponding to the minimum value of $\chi^2$ was adopted
as the starting guess for a further $\chi^2$ minimization based on the
downhill simplex method \citep{Nelder1965}. The resulting $\chi^2$
minimum, $\chi^2_{\rm min}$, corresponds to the best-fitting model of
the nuclear disc. 

We determined $\Delta \chi^2 = \chi^2 - \chi^2_{\rm min}$ and derived
its confidence levels under the assumption that the errors are
normally distributed and after rescaling $\chi_{\rm min}^2$ to be
equal to the number of degrees of freedom. It is $\chi_{\rm min}^2 =
N-4$, where $N$ is the number of fitted isophotes in the NSD region.
The best-fitting values and confidence levels of $\mu_0$, $h$, $i$,
and PA$_{\rm NSD}$ alone, marginalizing over all other parameters, are
shown in Fig.~\ref{fig:confidence}. The photometric parameters of the
NSD and their $3\sigma$ errors are listed in
Table~\ref{tab:parameters}.  The comparison between the isophotal
parameters of NGC~1023 measured before and after the subtraction of
the best-fitting model of the NSD is shown in
Fig.~\ref{fig:decomposition}.

\begin{table*}
\caption{Line-strength indices and stellar population properties in
 the concentric circular rings centred on the galaxy
 nucleus.}
\begin{tabular}{clccrrc}
\hline
\multicolumn{1}{c}{Id. No.} &
\multicolumn{1}{c}{Radius} &
\multicolumn{1}{c}{\Hb} &
\multicolumn{1}{c}{\MgFe} &
\multicolumn{1}{c}{$t$} &
\multicolumn{1}{c}{\FeH} &
\multicolumn{1}{c}{$L_{\rm NSD}/L_{\rm T}$} \\ 
\multicolumn{1}{c}{} &
\multicolumn{1}{c}{[arcsec]} &
\multicolumn{1}{c}{[\AA]} &
\multicolumn{1}{c}{[\AA]} &
\multicolumn{1}{c}{[Gyr]} &
\multicolumn{1}{c}{[dex]} &
\multicolumn{1}{c}{} \\
\multicolumn{1}{c}{(1)} &
\multicolumn{1}{c}{(2)} &
\multicolumn{1}{c}{(3)} &
\multicolumn{1}{c}{(4)} &
\multicolumn{1}{c}{(5)} &
\multicolumn{1}{c}{(6)} &
\multicolumn{1}{c}{(7)} \\
\hline
1 & 0.00-0.65 & $1.57\pm0.10$ & $3.99\pm0.23$ & $ 6.7\pm3.2$ & $ 0.43\pm0.25$ & 0.06\\
2 & 0.65-1.95 & $1.80\pm0.10$ & $3.85\pm0.21$ & $ 3.4\pm1.7$ & $ 0.56\pm0.24$ & 0.03\\
3 & 1.95-3.25 & $1.46\pm0.10$ & $3.57\pm0.20$ & $15.6\pm4.7$ & $ 0.01\pm0.17$ & 0.00\\
4 & 3.25-4.55 & $1.45\pm0.10$ & $3.42\pm0.19$ & $17.9\pm4.9$ & $-0.10\pm0.16$ & 0.00\\
5 & 4.55-5.85 & $1.43\pm0.10$ & $3.33\pm0.18$ & $19.7\pm4.2$ & $-0.16\pm0.12$ & 0.00\\
6 & 5.85-7.15 & $1.23\pm0.10$ & $3.22\pm0.18$ & ...          & ...            & 0.00\\
7 & 7.15-8.45 & $1.59\pm0.10$ & $3.03\pm0.17$ & $16.8\pm3.8$ & $-0.29\pm0.12$ & 0.00\\
\hline
\end{tabular}
\begin{minipage}{11cm}
{\em Note.} Column (1): identification number of the concentric
circular ring.  Column (2): minimum and maximum radius of the
concentric circular ring. Columns (3)-(4): equivalent widths of the
line-strength indices measured in the concentric ring. The values of
\Hb\ are directly from \citet{Silchenko1999}, while the values of
\MgFe\ are calculated from the data of \citet{Silchenko1999}. Columns
(5)-(6): age and iron abundance of the total stellar population based
on the SSP models of \citet{Worthey1994a}. Column (7): NSD-to-total
luminosity ratio within the ring in the WFPC2/{\em F555W}
passband.
\end{minipage}
\label{tab:indices}
\end{table*}

The location and orientation of the NSD (i.e. position of the centre,
inclination, and position angle) are the same in all the available
images. But, the NSD appears more concentrated in the bluer image,
given that the value for the scalelength obtained with the ACS/{\em
  F475W} image is inconsistent with those found for the WFPC2/{\em
  F555W} and {\em F814W} images even when considering 3$\sigma$
errors. This also implies the presence of colour gradients in the NSD,
an important constraint on the star formation process
\citep{Morelli2010}. The mean values of the NSD parameters are
$\langle h \rangle\,=\, 32.9$ pc, $\langle i \rangle\,=\, 77\fdg6$,
and $\langle {\rm PA} \rangle\,=\, 85\fdg7$. The size and luminosity
of the NSD of NGC~1023 are consistent with those of the other NSDs
detected so far \citep[see][for a census]{Ledo2010}.

\section{Stellar population properties of the nuclear disc}
\label{sec:population}

\subsection{Integral-field spectroscopy}
\label{sec:spectroscopy}

The integral-field spectroscopic observations of the nucleus of
NGC~1023 were carried out with the 6 m telescope of the Special
Astrophysical Observatory by \citet{Silchenko1999}.
The Multi-Pupil Field Spectrograph (MPFS) mapped a field of view of
$11\,\times\,21$ arcsec$^2$ at $\rm PA=122^\circ$ with $8\,\times\,16$
spectra. The spectral range between 4600 and 5450 \AA\ was covered
with an instrumental resolution of 5 \AA\ (FWHM) corresponding to
$\sigma_{\rm inst} = 131$ km s$^{-1}$ at \Hb. The exposure time on the
galaxy was 3600 s with an average seeing $\rm FWHM = 1.6$
arcsec. Further details about the observations and data reduction are
given in \citet{Silchenko1999}.

The resulting spectra were summed to obtain azimuthally averaged
spectra within concentric circular rings centred onto the galaxy
nucleus (Table~\ref{tab:indices}). This allowed to increase the
signal-to-noise ratio and derive more precise values for the \Mgb,
Fe5270, Fe5335, and \Hb\ Lick/IDS line-strength indices, as defined by
\citet{Worthey1994b}. The mean accuracy of the line-strength indices
measured in the azimuthally averaged spectra is 0.10
\AA\ \citep{Silchenko1999}.

We extracted the values of the \Mgb, Fe5270, \Fe, and \Hb\ indices
from figs. 4a and 5a in \citet{Silchenko1999}. Then, we calculated the
values of the Fe5335 index and derived the magnesium-iron index,
\MgFe$\,=\,\sqrt{{\rm Mg}b\;(0.72 \times {\rm Fe5270} + 0.28 \times
  {\rm Fe5335})}$ as defined by \citet{Thomas2003}. The equivalent
widths of the \Hb\ and \MgFe\ line-strength indices measured in the
azimuthally averaged spectra and their $1\sigma$ errors are listed in
Table~\ref{tab:indices}. The \Hb\ measurements by
\citet{Silchenko1999} are in agreement with those by
\citet{Kuntschner2006}, which are corrected for emission-line infill
and based on integral-field spectroscopic data obtained with the
Spectroscopic Areal Unit for Research of Optical Nebulae (SAURON). As
a matter of fact, no significant \Hb\ emission is detected within 2
arcsec from the centre, whereas it is very weak and patchy within 9
arcsec \citep{Sarzi2006}.

\begin{figure*}
\includegraphics[trim=25mm 15mm 15mm 20mm, width=16cm, angle=0, clip=yes]{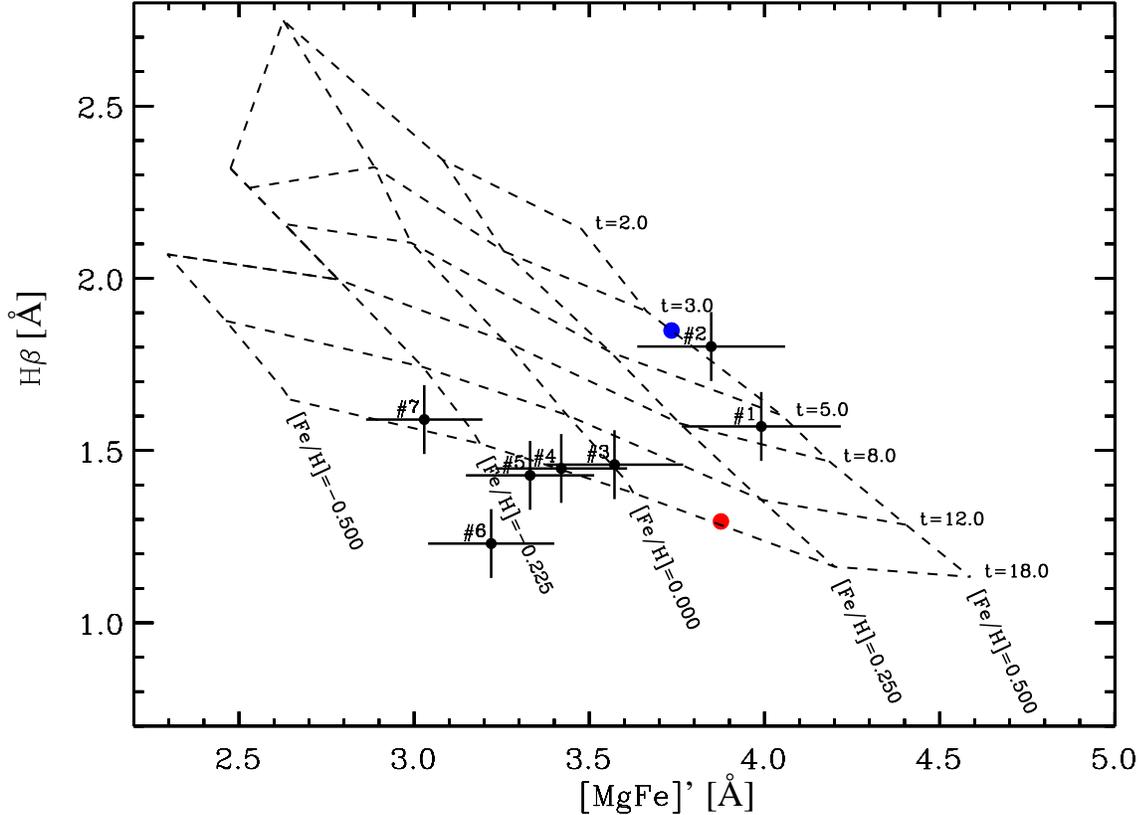}
\caption{Equivalent width of the \Hb\ and \MgFe\ line-strength indices
  measured in the azimuthally averaged spectra of NGC~1023 (black
  circles marked with identification numbers as in
  Table~\ref{tab:indices}) and predictions from SSP models with
  different ages (in Gyr) and iron abundances (in dex) by
  \citet[][dashed lines]{Worthey1994a}. The red filled circle and blue
  filled square correspond to the line-strength indices of the SSP
  adopted for the bulge ($t_{\rm bulge}=17.5$ Gyr, \FeH$_{\rm
    bulge}\,=\,0.13$ dex) and best-fitting SSP of NSD ($t_{\rm
    NSD}=3.4$ Gyr, \FeH$_{\rm NSD}\,=\,0.50$ dex) in the constrained
  fit, respectively (see Section~\ref{sec:csp}).}
\label{fig:ssp}
\end{figure*}

\subsection{Stellar population analysis}
\label{sec:analysis}

\subsubsection{Age and iron abundance of the total stellar population}
\label{sec:ssp}

We derived the radial trend of the properties of the total stellar
population using the evolutionary population models by
\citet{Worthey1994a}, which provide the values of the equivalent
widths of the Lick/IDS line-strength indices for a single stellar
population (SSP) of arbitrary age $t$ and iron abundance \FeH, given a
specific initial mass function\footnote{Model outputs also include
  magnitudes, colours, mass-to-light ratios, and spectral energy
  distribution of the SSPs and are available at {\tt
    http://astro.wsu.edu/dial/dial\_a\_model.html}.}. The models cover
the ages $1\,\leq\,t\,\leq\,18$ Gyr and iron abundances
$-2.0\,\leq\,[\ensuremath{{\rm Fe}/{\rm H}}]\,\leq 0.5$ dex, except
for $t\,\leq\,8$ Gyr and $[\ensuremath{{\rm Fe}/{\rm H}}]\,\leq
-0.225$ dex. We adopted the \citet{Salpeter1955} initial mass function
for our analysis.

Among the measured line-strength indices, we considered \Hb\ and \MgFe
. The former is sensitive to warm turn-off stars and, thus, it is an
useful age indicator. The latter provides an estimation of the total
metal abundance, almost independently of the \aFe\ ratio between the
$\alpha$ and iron-peak elements \citep{Thomas2003}. The equivalent
widths of \Hb\ and \MgFe\ measured in the azimuthally averaged spectra
of NGC~1023 are compared with the model predictions in
Fig.~\ref{fig:ssp}.
In order to derive the parameter pairs $(t,\,[{\rm Fe}/{\rm H}])$ from
the measured values (\Hb,\,\MgFe), we linearly interpolated between the 
points of the model grid.
We excluded the ring $5.85\,<\,R\,<\,7.15$ arcsec from the analysis
because its values of \Hb\ and \MgFe\ are not consistent with the
model predictions of age and iron abundance. 
The stellar-population age and iron abundance we calculated for the
remaining concentric rings centred on the nucleus of NGC~1023 are
given in Table~\ref{tab:indices}. They are consistent with the results
by \citet{McDermid2006} and \citet{Kuntschner2010}. 

The radial profiles of stellar age and iron abundance observed in
Fig.~\ref{fig:ssp} and tabulated in Table~\ref{tab:indices} are
plotted in Fig.~\ref{fig:gradients}. The inner regions of NGC~1023
($R\,<\,1.95$ arcsec) are younger ($t\,\approx\,5$ Gyr) and more metal
rich (\FeH$\,\approx\,0.5$ dex) than the outer regions
($1.95\,<\,R\,<\,8.45$ arcsec). In fact, at these larger radii we
measured old ages ($t\,\approx\,17$ Gyr) and a decrease of iron
abundance. It ranges from \FeH$\,\approx\,0$ dex to $-0.3$ dex with
increasing radius.  These are recurrent trends for the stellar
populations of S0 bulges \citep[e.g.][]{SanchezBlazquez2006,
  Morelli2008, Kuntschner2010}.

\subsubsection{Age and iron abundance of the bulge and nuclear disc}
\label{sec:csp}

\begin{figure}
\includegraphics[trim=15mm 70mm 25mm 25mm, width=8.0cm, angle=0, clip=yes]{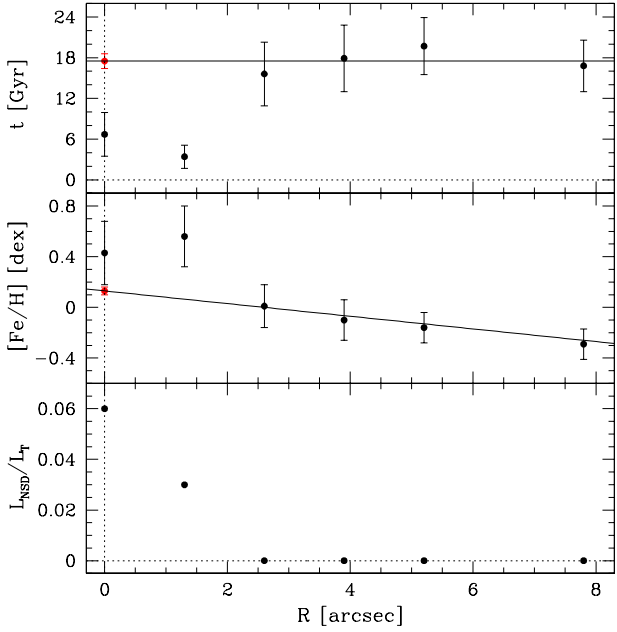}
\caption{The radial profile of age (top panel), iron abundance (middle
  panel), and NSD-to-total luminosity ratio (bottom panel) measured in
  the concentric circular rings centred on the galaxy nucleus (black
  circles). The solid lines fit a constant age (top panel) and
  linearly-increasing iron abundance (middle panel) in the outer rings
  ($R\,>\,1.95$ arcsec).  The red circles correspond to the age
  ($t_{\rm bulge}=17.5$ Gyr, top panel) and iron abundance (\FeH$_{\rm
    bulge}\,=\,0.13$ dex, middle panel) adopted for the SSP of the
  bulge in the constrained fit (see Section~\ref{sec:csp}).}
 \label{fig:gradients}
\end{figure}

The age and iron abundance variations of the total stellar population
in the centre of NGC~1023 (Fig.~\ref{fig:gradients}) can be explained
with the result of the superposition of two stellar populations with
different properties (i.e. the bulge and NSD stellar populations).

To constrain the age and iron abundance of the bulge and NSD, we
derived their light contribution to the total luminosity of the galaxy
in the concentric circular rings listed in Table~\ref{tab:indices}. We
analysed the WFPC2/{\em F555W} and disc model images after convolving
with a Gaussian PSF with $\rm FWHM = 1.6$ arcsec to match the
spectroscopic observations. The WFPC2/{\em F555W} passband was chosen
to cover more suitably the measured line-strength indices. We found
that the NSD contributes 6 and 3 per cent of the total luminosity in
the two innermost apertures ($R\,<\,1.95$ arcsec), whereas the bulge
contributes all the galaxy light in the remaining outer rings
($1.95\,<\,R\,<\,8.45$ arcsec, Table~\ref{tab:indices}). We did not
take into account the light contribution of the main stellar disc,
because it becomes significant only at much larger radii \citep[$R>70$
  arcsec;][]{Debattista2002, Noordermeer2008} than those we
considered.

Since $L_{\rm NSD}/L_{\rm T} = 0$ for $1.95\,<\,R\,<\,8.45$ arcsec, we
extrapolated at $R=0$ arcsec the values of age and iron abundance
derived for the outer rings in order to have the best possible model
of the stellar population of the bulge in the innermost circular
aperture.
We assumed the bulge age and its associated 1$\sigma$ uncertainty to
be equal to the mean age and standard deviation of the mean in the
outer rings, respectively. We found $t_{\rm bulge}\,=\,17.5\pm1.1$ Gyr
(Fig.~\ref{fig:gradients}).
We linearly interpolated the iron abundances in the outer rings to
estimate the iron abundance of the bulge and its 1$\sigma$ uncertainty
as the central iron abundance and its standard deviation,
respectively. To this aim, we rescaled the uncertainties on iron
abundance to have $\chi^2$ value equal to the number of degrees of
freedom for the best-fitting straight line.  We obtained \FeH$_{\rm
  bulge}\,=\,0.13\pm0.03$ dex (Fig.~\ref{fig:gradients}).

We exploited the inferred constraints on the NSD and bulge light
fractions and on the bulge stellar population in the innermost
circular aperture in order to look for the optimal combination of SSP
models of \citet{Worthey1994a} for bulge and NSD matching the measured
\Hb\ and \MgFe\ line-strength indices.

We calculated the total equivalent width of the line-strength index
$I$ (i.e. \Hb, \Mgb, Fe5270, and Fe5335) of the composite stellar
population of bulge and NSD as
%
\begin{equation}
 I_{\rm NSD+bulge} = \Delta\lambda
  \left(1 - \frac{F_{\rm line}^{\rm NSD+bulge}}{F_{\rm cont}^{\rm NSD+bulge}}\right)
\end{equation}
%
where $\Delta\lambda = 28.75, 32.5, 40.0,$ and $40.0$ \AA\ are the
widths of the line band for \Hb, \Mgb, Fe5270, and Fe5335,
respectively. The total continuum and line fluxes of the line-strength
index of the composite stellar population were derived as
%
\begin{eqnarray}
F_{\rm cont}^{\rm NSD+bulge} & = & w_{\rm NSD}
    \frac{F_{\rm cont}(t_{\rm NSD}, [{\rm Fe/H}]_{\rm NSD})}{
      \frac{M}{L}(t_{\rm NSD}, [{\rm Fe/H}]_{\rm NSD})} + \nonumber \\
                        &   & w_{\rm bulge}
    \frac{F_{\rm cont}(t_{\rm bulge}, [{\rm Fe/H}]_{\rm bulge})}{
      \frac{M}{L}(t_{\rm bulge}, [{\rm Fe/H}]_{\rm bulge})}, 
\end{eqnarray}
%
and 
%
\begin{eqnarray}
F_{\rm line}^{\rm NSD+bulge} & = & w_{\rm NSD}
    \frac{F_{\rm line}(t_{\rm NSD}, [{\rm Fe/H}]_{\rm NSD})}{
      \frac{M}{L}(t_{\rm NSD}, [{\rm Fe/H}]_{\rm NSD})} + \nonumber \\
                        &   & w_{\rm bulge}
    \frac{F_{\rm line}(t_{\rm bulge}, [{\rm Fe/H}]_{\rm bulge})}{
      \frac{M}{L}(t_{\rm bulge}, [{\rm Fe/H}]_{\rm bulge})}, 
\end{eqnarray}
%
respectively. The mass-weighted line and continuum fluxes of the
line-strength index, $F_{\rm line}(t, [{\rm Fe/H}])$ and $F_{\rm
  cont}(t, [{\rm Fe/H}])$, and the mass-to-light ratio, $M/L(t, [{\rm
    Fe/H}])$, of the SSPs of bulge and NSD as a function of age and
iron abundance were taken from the same models of \citet{Worthey1994a}
used in Section~\ref{sec:ssp}. In addition, we adopted the weights
$w_{\rm NSD} = 0.06$ and $w_{\rm bulge} = 0.94$ to account for the
light contribution of the two components in the innermost circular 
aperture of NGC~1023. This allowed us to derive the equivalent widths
of \Hb\ and \MgFe\ for any composite stellar population of bulge and
NSD as soon as we defined their ages $t_{\rm bulge}$ and $t_{\rm
  NSD}$, and iron abundances $[{\rm Fe/H}]_{\rm bulge}$ and $[{\rm
    Fe/H}]_{\rm NSD}$.

We determined the best-fitting age and iron abundance for bulge and
NSD in two different ways by minimizing the $\chi^2$ function defined
as
\begin{equation}
\chi^2 = 
  \frac{({\rm H}\beta - {\rm H}\beta_{\rm NSD+bulge})^2}{\sigma_{{\rm H}\beta}^2} +
  \frac{([{\rm MgFe}]' - [{\rm MgFe}]'_{\rm NSD+bulge})^2}{{\sigma_{[{\rm MgFe}]'}^2}}
\end{equation}
where $\sigma_{{\rm H}\beta}$ and $\sigma_{[{\rm MgFe}]'}$ are the
errors on the equivalent widths of \Hb\ and \MgFe\ measured in the
innermost circular aperture.

\begin{figure*}
\includegraphics[trim=15mm 0mm 45mm 150mm, width=16cm, angle=0, clip=yes]{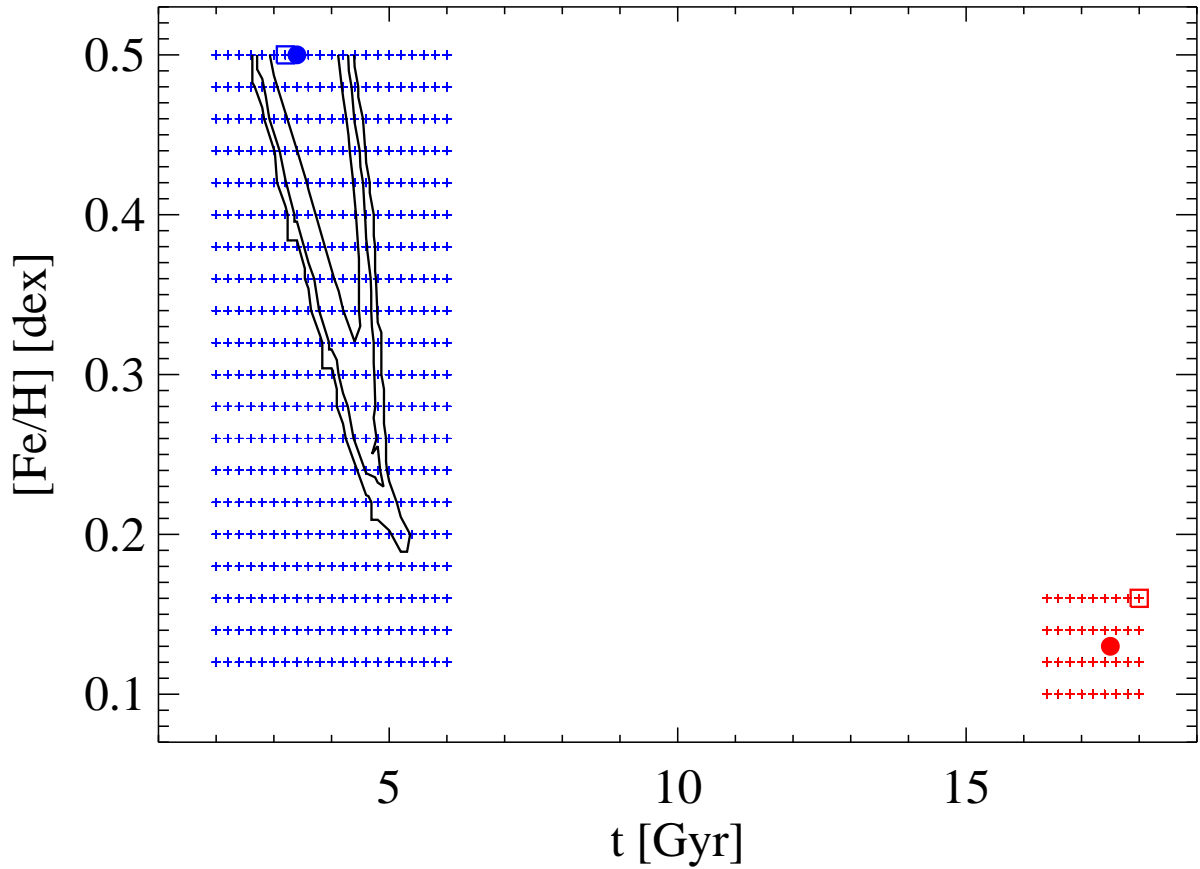}
\caption{SSP models of the bulge (red times) and NSD (blue crosses)
  as a function of age and metallicity from \citet{Worthey1994a} used
  to build the composite stellar population with predicted values of
  the \Hb\ and \MgFe\ line-strength indices matching those measured in
  the innermost circular aperture of NGC~1023. The symbols mark the
  SSP adopted for the bulge ($t_{\rm bulge}=17.5$ Gyr, \FeH$_{\rm
    bulge}\,=\,0.13$ dex; red filled circle) and best-fitting SSP of
  the NSD ($t_{\rm NSD}=3.4$ Gyr, \FeH$_{\rm NSD}\,=\,0.50$ dex; blue
  filled square) in the constrained fit and the best-fitting SSPs of
  the bulge ($t_{\rm bulge}=18.0$ Gyr, \FeH$_{\rm bulge}\,=\,0.16$
  dex; red open circle) and NSD ($t_{\rm NSD}=3.2$ Gyr, \FeH$_{\rm
    NSD}\,=\,0.50$ dex; blue open square) in the unconstrained
  fit. The contours encompass the 68.3, 95.4, and 99.7 per cent of the
  models of composite stellar population with predicted values of
  \Hb\ and \MgFe\ line-strength indices consistent with the measured
  ones within the errors, respectively.}
\label{fig:contours}
\end{figure*}

In the first approach, we performed a constrained fit by adopting for
the bulge $t_{\rm bulge}\,=\,17.5$ Gyr and \FeH$_{\rm bulge}\,=\,0.13$
dex. We looked for the NSD properties on a fine grid in age from 2 to
6 Gyr (on steps of 0.2 Gyr) and iron abundance from 0.12 to 0.50 dex (on
steps of 0.02 dex) obtained by linearly interpolating the models of
\citet{Worthey1994a}. We found that the best-fitting model of NSD
stellar population has $t_{\rm NSD}\,=\,3.4$ Gyr and \FeH$_{\rm
  NSD}\,=\,0.50$ dex (Fig.~\ref{fig:contours}).

In the second approach, as a consistency check, we performed an
unconstrained fit by allowing for the bulge age to vary between 16.4
and 18.0 Gyr (on steps of 0.2 Gyr) and bulge iron abundance to vary
between 0.10 and 0.16 dex (on steps of 0.02 dex) to account for the
adopted uncertainties on the bulge properties. For the NSD, we explored
the same ranges of age and iron abundance as for the constrained
fit. The resulting best-fitting model requires a bulge with $t_{\rm
  bulge}\,=\,18.0$ Gyr and \FeH$_{\rm bulge}\,=\,0.16$ dex and an NSD
with $t_{\rm NSD}\,=\,3.2$ Gyr and \FeH$_{\rm NSD}\,=\,0.50$ dex
(Fig.~\ref{fig:contours}).

In both cases, we conclude that the NSD of NGC~1023 is much younger
and more metal rich than the host bulge. The age of the NSD is well
constrained, whereas the its iron abundance corresponds to the upper
limit of the \FeH\ range of the available SSP models. Without the
knowledge of the light fraction of NSD, the age and metallicity of the
NSD can be hardly recovered due to the degeneracy between these two
parameters \citep[see][for a discussion]{Sarzi2015b}.

\section{Discussion and conclusions}
\label{sec:discussion}

The structure and properties of the stellar population of the nuclear
region of the interacting SB0 galaxy NGC~1023 were investigated
through a detailed analysis of archival {\em HST} imaging and
ground-based integral-field spectroscopy.
The galaxy is known to host a highly-inclined NSD \citep{Faber1997,
  Silchenko1999, Sarzi2006, Ledo2010}. Moreover, the stars of the
galaxy nucleus have significantly different chemical properties with
respect to the surrounding bulge. They are younger, more metal rich,
and have a higher magnesium overabundance \citep{Silchenko1999,
  McDermid2006, Kuntschner2010}.

In this paper, we derived the central surface brightness, scalelength,
inclination, and position angle of the NSD of NGC~1023 in all the
available broad band {\em HST} images by assuming that it is an
infinitesimally thin exponential disc. We applied the photometric
decomposition method of \citet{Scorza1995} and adopted the algorithm
by \citet{Morelli2004, Morelli2010}, which we implemented for a better
determination of the best-fitting parameters. The location and
orientation of the NSD do not depend on the observed passband, as
already observed for the few other NSDs for which a detailed multiband
photometric analysis was performed \citep{Krajnovic2004, Morelli2010,
  Corsini2012} whereas the NSD appears more concentrated in the bluer
image.
We measured the light contribution of the NSD to the total
luminosity of the galaxy $L_{\rm NSD}/L_{\rm T}$ in the same
concentric circular rings centred on the galaxy nucleus where the
equivalent widths of the Lick/IDS line-strength indices were obtained
from ground-based integral-field spectroscopy by
\citet{Silchenko1999}.  In the WFPC2/{\em F555W} passband the NSD
contributes a maximum of 6 per cent of the total luminosity within the
innermost circular aperture ($R < 0.65$ arcsec), whereas all the
galaxy light is given by the bulge in the outer rings ($1.95 < R <
8.45$ arcsec).

The ratio $L_{\rm NSD}/L_{\rm T}$ was used in combination with the SSP models by
\citet{Worthey1994a} to derive the properties of the stellar
population of the NSD by matching the measured \Hb\ and
\MgFe\ line-strength indices.  For the host bulge we assumed an old
stellar population ($t\,=\,14.3$ Gyr) with super-solar iron abundance
(\FeH$\,=\,0.13$ dex) to be consistent with the constant age and
gradient of iron abundance which we measured in the outer rings where
the light contribution of the NSD is null. We found that the nuclear
disc of NGC~1023 is much younger ($t\,=\,3.4$ Gyr) and more metal rich
(\FeH$\,=\,0.50$ dex) than the host bulge.

These new results about NGC~1023 give us the opportunity to gain a
further insight on the formation of NSDs and their host
galaxies. Indeed, the only other NSD, whose stellar population was
disentangled from that of its host galaxy is that in NGC~4458
\citep{Sarzi2015b}. NGC~4458 is an intermediate-mass elliptical in the
Virgo cluster. Its NSD is older than 6 Gyr, implying that the galaxy
did not experience a disruptive event since $z\sim0.6$. Therefore,
NGC~4458 assembled most of its mass long time ago, consistently with
the picture where smaller galaxies formed earlier and high-density
environments reduce the incidence of mergers.
On the contrary, the properties of the stellar populations of the other NSDs
studied so far were derived by assuming a negligible contamination of
the host galaxy to the colours and line-strength indices measured in
nucleus. According to this assumption, some NSDs are found to be as
old as the host galaxy (NGC~4128, \citealt{Krajnovic2004}; NGC~4342,
\citealt{vandenBosch_etal1998}; NGC~4458, \citealt{Morelli2004};
NGC~4621, \citealt{Krajnovic2004}; NGC~4698,
\citealt{Corsini2012}). The NSD of NGC~4570 shows the same
intermediate age, but it is more metal rich than the galaxy bulge
\citep{Krajnovic2004}. In NGC~4478 the NSD is younger, more metal
rich, and it has a lower magnesium overabundance than the main body of
the galaxy \citep{Morelli2004}. In contrast, the NSD of NGC~5308 is
made of a younger and more metal-poor stellar population than the host
galaxy \citep{Krajnovic2004}. Finally, a few cases are characterised
by on-going star formation (NGC~5845, \citealt{Kormendy1994};
NGC~4486A, \citealt{Kormendy2005}). 
 
As far as NGC~1023 concerns, the younger age of its NSD excludes a
scenario in which it assembled from already formed stars, accreted
from star clusters that migrated from the much older bulge
\citep[e.g.][]{Antonini2012, Portaluri2013}. This points toward a
dissipational process and in-situ star formation of gas funnelled into
the galaxy centre. The stars of the NSD have super-solar iron
abundance. Therefore, we argue they formed from processed gas, which
can be of either internal or external origin. The main disc of
NGC~1023 and interacting satellite NGC~1023A are both a possible
reservoir of processed gas.  The mild kinematic misalignment between
stars and ionized gas measured in the velocity fields provided by
SAURON is suggestive of a gradual change of the gas angular momentum
as a function of radius \citep{Sarzi2006}. This could be interpreted
as the result of the on-going interaction suffered by NGC~1023 and
unveiled by the complex distribution of the neutral hydrogen.
Processed gas could have been also accreted by NGC~1023 during the
merger with a small companion which occurred $\sim2$ Gyr ago,
according to the results of the smooth particle hydrodynamical
simulations performed by \citet{Bettoni2012}. However, NGC~1023 has a
bar which may have efficiently driven the gas to the galaxy centre,
where it settled into a nuclear disc and fuelled star formation. Such
a mechanism has been recently investigated by \citet{Cole2014}
studying the bar-induced gas inflows in a disc galaxy using a smooth
particle hydrodynamical simulation at high resolution. The simulation
confirms that gas dissipation is a key ingredient for forming a thin,
kinematically cool, young, and metal-rich structure within the central
500 pc of a galaxy. Assessing the properties of the stellar
populations in the main disc of NGC~1023 and measuring the metallicity
of the gas in the stream arising from the on-going merger between
NGC~1023 and NGC~1023A will allow a direct test of any role they may
have played in the formation of the NSD.

In this respect, our understanding of NSDs would benefit from an
integral-field spectroscopic follow up of a representative sample of
NSD galaxies. The optimized extraction of the spectrum of the host
spheroid in combination with the analysis of spectra obtained in
carefully selected regions, where the light contribution of the NSD is
maximal, will give tighter constraints on the disc stellar population
and unveil whether NGC~1023 and NGC~4458 are harbouring typical or
peculiar NSDs.

\section*{Acknowledgments}

We thank the anonymous referee for the valuable suggestions that
improved this manuscript.  We are especially grateful to Lodovico
Coccato, Marc Sarzi, and Olga Sil'chenko for insightful discussions on
the properties of nuclear stellar discs and to Claudia Maraston,
Alexadre Vazdekis, and Guy Worthey for very helpful comments on the
stellar population analysis.  This work is supported by Padua
University through grants 60A02-5857/13, 60A02-5833/14, 60A02-4434/15,
and CPDA133894. L.M. acknowledges financial support from Padua
University grant CPS0204. E.P. is partially supported by Fondazione
Angelo Della Riccia and acknowledges the Jeremiah Horrocks Institute
of the University of Central Lancashire for the hospitality while this
paper was in progress.

\label{lastpage}

\end{document}